\documentclass[9pt,conference]{IEEEtran} %updated techreport 2016-02-11
   \usepackage{times}
   \usepackage{mathptm}
   \usepackage{caption}
   \usepackage{epsfig}
    \usepackage{subfigure}
   \usepackage{bbold}
   \usepackage{color}
   \usepackage{latexsym}
   \usepackage{pslatex}
   \usepackage{algorithmic}
   \usepackage{algorithm}
   \usepackage{graphicx}
  \usepackage{multirow}

\newtheorem{definition}{Definition}

\newtheorem{theorem}{Theorem}

\begin{document}

\title{On Constructing Secure and Hardware-Efficient Invertible Mappings}
% over $GF(2^n)$.}
\author{
Elena~Dubrova  \\
Royal Institute of Technology (KTH), Stockholm, Sweden \\ 
dubrova@kth.se
}

\maketitle
 
%------------------------------------------------------------------------- 
\begin{abstract}
Our society becomes increasingly dependent on wireless communications. The tremendous growth in the number and type of wirelessly connected devices in a combination with the dropping cost for performing cyberattacks create new challenges for assuring security of services and applications provided by the next generation of wireless communication networks. The situation is complicated even further by the fact that many end-point Internet of Things (IoT) devices have very limited resources for implementing security functionality. This paper addresses one of the aspects of this important, many-faceted problem - the design of hardware-efficient cryptographic primitives suitable for the protection of resource-constrained IoT devices. We focus on cryptographic primitives based on the invertible mappings of type $\{0,1,\ldots,2^n-1\} \rightarrow \{0,1,\ldots,2^n-1\}$.
In order to check if a given mapping is invertible or not, we generally need an exponential in $n$ number of steps.
In this paper, we derive a sufficient condition for invertibility which can be checked in $O(n^2 N)$ time, where $N$ is the size of representation of the largest function in the mapping. Our results can be used for constructing cryptographically secure invertible mappings which can be efficiently implemented in hardware.
\end{abstract}

%-----------------------------------------------------------------------
%------------------------------------------------------------------------- 

\section{Introduction}

This paper addresses the problem of constructing hardware-efficient cryptographic primitives 
suitable for assuring trustworthiness of resource-constrained devices used in services and applications provided by the next generation of wireless communication networks.

The importance of improving security of wireless networks for our society is hard to overestimate. 
In 2014, the annual loss to the global economy from cybercrimes was more than \$400 billion~\cite{Cy14}.
This number can  quickly grow larger with the rapid growth of {\em Internet-of-Things} (IoT) applications. 
In coming years "things" such as household appliances, meters, sensors, vehicles, etc. are expected to be accessible and controlled via local networks or the Internet, opening an entirely new range of services appealing to users.  
The ideas of self-driving cars, health-tracking wearables, and remote surgeries do not sound like science-fiction any longer. The number of wirelessly connected devices is  expected
to grow to a few tens of billions in the next five years~\cite{Er12}.

Unfortunately, the new technologies are appealing to the attackers as well. As processing power and connectivity become
cheaper, the cost of performing a cyberattack drops, making it easier for adversaries of all types to penetrate networks. Attacks are becoming more frequent, more sophisticated, and more widespread.
A connected household appliance becomes a target for all hackers around the globe unless appropriate security mechanisms are implemented. Household  appliances typically do not have the same level of protection as computer systems. A compromised device can potentially be used as an entry point for a cyberattack on other devices connected to the network.  The first proven cyberattack involving "smart" household  appliances has been already reported in~\cite{Pr14}. In this attack, more than 750.000 malicious emails targeting enterprises and individuals worldwide were sent from more than 100.000 consumer devices such as home-networking routers, multi-media centers, TVs, refrigerators, etc. No more than 10 emails were initiated from any single IP-address, making the attack difficult to block based on location. The attack surface of future IoT with billions of connected devices
will be enormous.

In addition to a larger attack surface, the return value for performing a cyberattack grows. 
The assets accessible via tomorrow's networks (hardware, software, information, and revenue streams) are expected to be much greater than the ones available today,  
increasing incentive for cyber criminals and underground economies~\cite{Er5G}. A growing black market for breached data serves as a further encouragement. 
The damage caused by an individual actor may not be limited to a business or reputation, but could have a severe impact on public safety, national economy, and/or national security. 

The tremendous growth in the number and type of wirelessly connected devices, as well as the increased incentive for performing attacks change the threat landscape and create new challenges for security. 
The situation is complicated even further by the fact that many 
end-point IoT devices require utmost efficiency in the use of communication, computing, storage and energy resources. 
A typical IoT device spends most of its "life" in a sleep mode. It gets activated at periodic intervals, transmits a small amount of data and then shuts down again. To satisfy extreme limitations 
of resource-constrained IoT devices, very efficient cryptographic primitives for implementing encryption, data integrity protection, authentication, etc. are required. 

Invertible mappings are among the most frequently used primitives in cryptography~\cite{St06}.
For example, the round function of a block cipher~\cite{pres07,pri12} has to be invertible in order
to result in unique decryption. Stream ciphers~\cite{hell-grain,DuH14} and hash functions~\cite{Quark,DuNS15} use invertible 
state mappings to prevent incremental reduction of the entropy of the state.  

This paper presents a sufficient condition for invertibility of mappings $x \rightarrow f(x)$  of type $\{0,1,\ldots,2^n-1\} \rightarrow \{0,1,\ldots,2^n-1\}$. Such a single-variable $2^n$-valued mapping can be interpreted as an $n$-variable Boolean mapping $\{0,1\}^n \rightarrow \{0,1\}^n$  in which the Boolean variable $x_i \in \{0,1\}$ represents the bit number $i$ of the input $x$
and the Boolean function $f_i: \{0,1\}^n \rightarrow \{0,1\}$ represents the bit number $i$ of the output $f(x)$, i.e.:
\begin{equation} \label{map}
\left(
\begin{array}{c}
x_0\\
\ldots \\
x_{n-1}\\
\end{array}
\right)
\rightarrow
\left(
\begin{array}{cc}
f_0(x_0,\ldots,x_{n-1})\\
\ldots \\
f_{n-1}(x_0,\ldots,x_{n-1})\\
\end{array}
\right)
\end{equation}
for $i \in \{0,1,\ldots,n-1\}$.  
For example, the 4-variable Boolean mapping
\[
\left(
\begin{array}{c}
x_0\\
x_1\\
x_2\\
x_3\\
\end{array}
\right)
\rightarrow
\left(
\begin{array}{c}
x_0 \oplus 1\\
x_1 \oplus x_0\\
x_2 \oplus x_0 x_1\\
x_3 \oplus x_0 x_1 x_2\\
\end{array}
\right)
\]
corresponds to the single-variable 16-valued mapping 
\[
x \rightarrow x + 1 ~~\mbox{(mod 16)},
\]
where ``$+$'' is addition modulo 16.
Note that the corresponding single-variable $2^n$-valued mapping may not have a closed form.

In order to check if a mapping of type~(\ref{map}) is invertible or not, we generally need an exponential in $n$ number of steps. The condition derived in this paper can be checked in $O(n^2 N)$ time,
where $N$ is the size of representation of the largest Boolean function in the mapping.
So, it can be used for constructing secure invertible mappings
with a small hardware implementation cost. For example, we show how the presented results can be used in
stream cipher design.

The paper is organized as follows. 
Section~\ref{not} summarizes basic notations used in the sequel.
In Section~\ref{prev}, we describe previous work and
show that previous methods cannot explain invertibility of some mappings which
can be handled by the presented approach.
Section~\ref{main} presents the main result. Section~\ref{compl}
estimates the complexity of checking the presented condition.
Section~\ref{appl} shows how the presented results can be used in
stream cipher design.
Section~\ref{conc} concludes the paper.

\section{Preliminaries} \label{not}

Throughout the paper, we use "$\oplus$" to 
denote the Boolean XOR, "$\cdot$"  
to denote the Boolean AND and $\overline{x}$ to denote the Boolean complement of $x$,
defined by $\overline{x} = 1 \oplus x$.  

The Algebraic Normal Form (ANF)~\cite{CuS09} of a Boolean function $f: \{0,1\}^{n} \rightarrow \{0,1\}$
(also called positive polarity {\em Reed-Muller canonical form}~\cite{Gr91}) is a polynomial in the Galois Field of order 2, $GF(2)$, of type
\[
f(x_0, x_1,\ldots,x_{n-1}) = \sum_{i=0}^{2^n-1}  c_i \cdot 
x_0^{i_0} \cdot x_1^{i_1} \cdot \ldots \cdot x_{n-1}^{i_{n-1}},
\]
where $c_{i} \in \{0,1\}$ are constants, ``$\cdot$'' is the Boolean AND, ``$\sum$'' is the Boolean XOR, the vector $(i_{0} i_{1} \ldots i_{n-1})$ is the binary expansion of $i$, and $x^{i_{j}}_{j}$ denotes the $i_{j}$th power of $x_{j}$ defined by
$x^{i_{j}}_{j} = x_{j}$ for $i_{j} = 1$ and $x^{i_{j}}_{j} = 1$ otherwise, for $j \in \{0,1, \ldots, n-1\}$.

The {\em dependence set} (or {\em support set}~\cite{espr}) of a Boolean function $f: \{0,1\}^n \rightarrow \{0,1\}$ 
is defined by
\[
dep(f) = \{j \ | \ f|_{x_j=0} \not = f|_{x_j=1}\},
\]
where $f|_{x_j=k} = f(x_0, \ldots, x_{j-1}, k, x_{j+1}, \ldots, x_{n-1})$
for $k \in \{0,1\}$.

A mapping $x \rightarrow f(x)$ on a finite set is called {\em invertible} if $f(x) = f(y)$ 
if, and only if, $x = y$. Invertible mappings are also called {\em permutations}.

\section{Previous Work} \label{prev}

In this section, we describe previous work on invertible mappings and
show that previous methods cannot explain invertibility of some mappings which
can be handled by the presented approach.

Many methods for constructing different classes of invertible mappings are known.
The simplest one is to compose simple invertible operations. A Substitution-Permutation Network (SPN)
is a typical example. An SPN consists of S-boxes, which permute input bits locally,
and P-boxes, which diffuse input bits globally. This method of construction
cannot explain the invertibility of, for example, the following 4-variable 
mapping
\begin{equation} \label{ex}
\left(
\begin{array}{c}
x_0\\
x_1\\
x_2\\
x_3\\
\end{array}
\right)
\rightarrow
\left(
\begin{array}{c}
x_1\\
x_2\\
x_3 \oplus x_1 x_2\\
x_0 \oplus x_3\\
\end{array}
\right)
\end{equation}
since a non-invertible operation Boolean AND is used.

Feistel~\cite{Fe73} proposed a powerful technique which makes possible constructing invertible mappings from non-invertible basic operations. It is used in many block ciphers, including DES~\cite{DES77}.
The basic Feistel construction maps two inputs, $l$ and $r$, into two outputs as follows:
\[
\left(
\begin{array}{c}
l\\
r\\
\end{array}
\right)
\rightarrow
\left(
\begin{array}{cc}
r\\
l \oplus f(r)\\
\end{array}
\right)
\]
where $f$ is any single-variable function. The full Feistel construction iterates the above mapping any number of times. This method was extended in several directions, including {\em unbalanced}, {\em homogeneous}, {\em heterogeneous}, {\em incomplete}, and {\em inconsistent} Feistel networks~\cite{ScK96}. However, the Feistel construction requires at least two variables. It cannot explain the invertibility of mappings $x \rightarrow f(x)$ of type $\{0,1,\ldots,2^n-1\} \rightarrow \{0,1,\ldots,2^n-1\}$. 
The presented method can explain it by looking into the structure of
Boolean functions $f_i$ representing the bits of the output $f(x)$.

Shamir~\cite{Sh93} introduced an interesting construction based on the fact that the mapping of type~(\ref{map}) 
is invertible for almost all inputs if each $f_i$ has the form:
\[
f_i(x_0,\ldots,x_{n-1}) = h_i(x_0,\ldots,x_{i-1}) x_i + g_i(x_0,\ldots,x_{i-1}) ~~~ (\mbox{mod} ~~ N)
\]
where $g_i$ and $h_i$ are arbitrary non-zero $i$-variable polynomials modulo a large RSA modulus $N$.
The "triangular" nature of functions $f_i$ makes it possible to perform
the inversion process similarly to the way we do Gaussian elimination
to solve a system of linear equations. 
This approach can also handle 
$g_i$'s and $h_i$'s which mix arithmetic and Boolean operations~\cite{KlS02,Kl05}. The approach presented in this paper is similar to the approach of Shamir in that it also uses triangulation to prove invertibility. However, our functions $f_i$ are of the type
\begin{equation} \label{el_eq}
f_i(x_0,\ldots,x_{n-1}) = x_j \oplus g_i(x_0,\ldots,x_{n-1}),
\end{equation}
where $j \in \{0,1,\ldots,n-1\}$ and $g_i$ does not depend on $x_j$.
Thus, in our case, the $i$th output may depend on the input $j$ such that $j > i$.
For example, in the mapping defined by equations~(\ref{ex}), 
$f_2$ depends on $x_1,x_2$ and $x_3$.

Shamir's construction was further extended by Klimov and Shamir~\cite{KlS02,Kl05} to a class on invertible mappings based on T-functions. A single-variable function $f(x)$ of type 
$\{0,1,$ $\ldots,2^n-1\} \rightarrow \{0,1,\ldots,2^n-1\}$ is defined to be a T-function
if each of its output bits $f_{i-1}(x)$ depends only on 
the first $i$ input bits $x_0, x_1, \ldots, x_{i-1}$:
\[
\left(
\begin{array}{l}
{x}_0 \\
{x}_1 \\
{x}_2 \\
\ldots \\
{x}_{n-1} \\
\end{array}
\right)
\rightarrow
\left(
\begin{array}{l}
f_0(x_0) \\
f_1(x_0, x_1) \\
f_2(x_0, x_1, x_2) \\
\ldots \\
f_{n-1}(x_0, x_1, \ldots, x_{n-1}) \\
\end{array}
\right).
\]
A T-function is invertible if, and only if, each output bit $f_i$ 
can be represented as 
\[
f_i(x_0,\ldots,x_i) = x_i \oplus g_i(x_0,\ldots,x_{i-1}).
\]
This fundamental result inspired the construction which we present in this paper.
The reader will easily notice that, in our construction~(\ref{el_eq}), 
functions $g_i$'s are T-functions while functions $f_i$'s are not if $j > i$.
Another difference is that Klimov and Shamir targeted software implementation
and therefore focused on mappings 
whose expression can be evaluated by a program with the minimal number 
of instructions. We target the hardware implementation.
Therefore, we are interested in minimizing the number of Boolean operations 
in the expressions of Boolean functions representing output bits.
The construction method which we present can be used as a starting point for
constructing nonlinear invertible mappings which have an efficient hardware implementation.

Another group of construction methods consider {\em permutation polynomials} which 
involve only the arithmetic operation of addition, subtraction, and multiplication.
Permutation polynomials are well-studied in mathematics. Hermite~\cite{LiH94} made a substantial progress in characterizing univariate permutation polynomials modulo a prime $p$. 
Dickson~\cite{Di01} described all univariate polynomials with degrees smaller than 5. However, the problem remains unsolved for high degree polynomials modulo a large prime $p$. 

The problem appears simpler for the ring of integers modulo $2^n$. Rivest~\cite{Ri99} provided a complete characterization of all univariate permutation polynomials modulo $2^n$. He proved that a polynomial 
\[
p(x) = a_0 + a_1 x + \ldots + a_d x^d
\]
with integral coefficients is a permutation polynomial for $n > 2$ if, and only if, $a_1$ is odd, $(a_2 + a_4 + \ldots)$ is even and $(a_3 + a_5 + \ldots)$ is even. 
His powerful algebraic proof technique was further generalized by Klimov and Shamir~\cite{KlS02,Kl05} to polynomials of type 
\[
p(x) = a_0 \circ a_1 x \circ \ldots \circ a_d x^d,
\]
where $\circ  \in \{+,-,\oplus\}$, which mix arithmetic and Boolean operations. However, since the resulting polynomials are T-functions, they do not cover the construction method presented in this paper for the case when $i$th output depends on the input $j$ such that $j > i$.

\begin{figure}[t!]
\begin{center}
\includegraphics[width=2.3in]{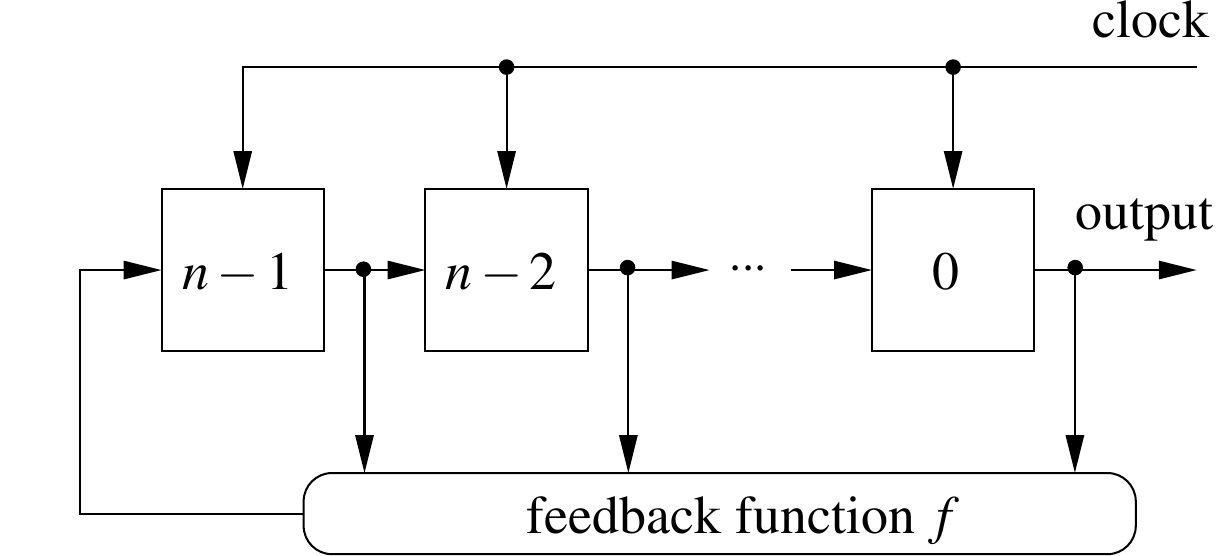}
\caption{The structure of an $n$-bit Non-Linear Feedback Shift Register.}\label{f1} 
\end{center}
\end{figure}

Finally, we would like to discuss the relation between the mappings of type~(\ref{map}) and
the state mappings generated by {\em Non-Linear Feedback Shift Registers (NLFSRs)}~\cite{Ja89}.
An $n$-bit NLFSR consists of $n$ binary stages, each capable of storing 1 bit of information, a nonlinear Boolean function, called {\em feedback function}, and a clock (see Figure~\ref{f1}). 
At each clock cycle, the  stage $n-1$ is updated to the value computed by the feedback function. The rest of the stages shift the content of the previous stage. Thus, an $n$-bit NLFSR with the feedback functions $f$ implements the state mapping of type 
\[
\left(
\begin{array}{c}
x_0\\
x_1\\
\ldots \\
x_{n-1}\\
\end{array}
\right)
\rightarrow
\left(
\begin{array}{cc}
x_1\\
x_2\\
\ldots \\
f(x_0,\ldots,x_{n-1})\\
\end{array}
\right).
\]
where the variable $x_i$ represents the value of the stage $i$, for $i \in \{0,1,\ldots,n-1\}$.

It is well-known~\cite{Golomb_book} that an $n$-bit NLFSR is invertible if, and only if, its feedback function is of type
\[
f(x_0,\ldots,x_{n-1}) = x_0 \oplus g(x_1,x_2,\ldots,x_{n-1}).
\]

\begin{figure}[t!]
\begin{center}
\includegraphics[width=3.4in]{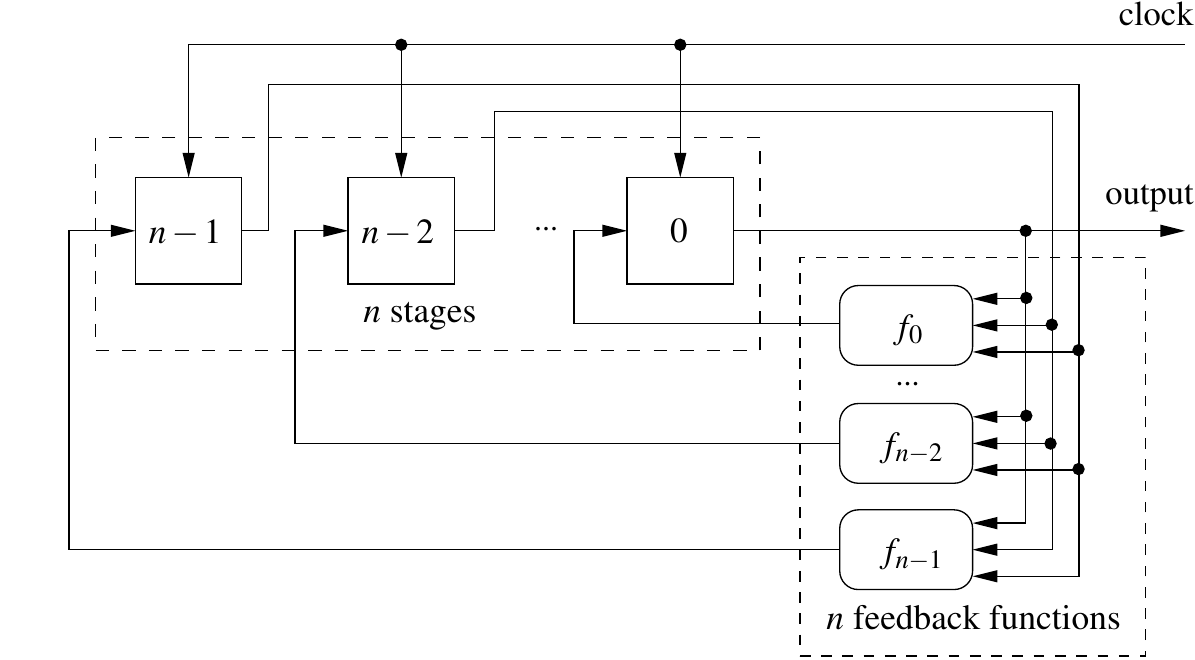}
\caption{The structure of an $n$-bit binary machine.}\label{f2} 
\end{center}
\end{figure}

The mappings considered in this paper can be implemented by a more general type of non-linear state machines, shown in Figure~\ref{f2}. Since the content of stages is not any longer shifted from one stage to the next, but rather it is updated by some arbitrary functions, 
such registers are not called shift registers any longer. Instead, they are called
{\em binary machines}~\cite{Golomb_book} or
{\em registers with non-linear update}~\cite{LiD14}.
Binary machines 
are typically smaller and
faster than NLFSRs generating the same sequence~\cite{Du10aj,Du11a}.
For example, the 4-bit NLFSR with the feedback function $f(x_0,x_1,x_2,x_3) = x_0 \oplus x_3 \oplus x_1 x_2 \oplus x_2 x_3$ generates the same set of sequences as the 4-bit binary machine implementing the 4-variable mapping defined by equations~(\ref{ex}). We can see that the binary machine uses 3 binary Boolean operations, while the NLFSR uses 5 binary Boolean operations. Furthermore, the depth of feedback functions of the binary machine is smaller that the depth of the feedback function of the NLFSR. Thus, the binary machine has a smaller propagation delay than the NLFSR.

%For example, consider the 4-bit NLFSR with the feedback function 
%\[
%f(x_0,x_1,x_2,x_3) = x_0 \oplus x_3 \oplus x_1 x_2 \oplus x_2 x_3.
%\]
%If this NLFSR is initialized to the state $(x_0 x_1 x_2 x_3) = (1000)$, it generates the output sequence 
%\[
%(1,0,0,0,1,1,0,1,0,1,1,1,1,0,0)
%\]
%with the period 15. The same sequence can be generated by the 4-bit binary machine implementing the 4-variable mapping defined by equations~(\ref{ex}). We can see that the binary machine uses 3 binary Boolean operations, while the NLFSR uses 5 binary Boolean operations. Furthermore, the depth of feedback functions of the binary machine is smaller that the depth of the feedback function of the NLFSR. Thus, the binary machine has a smaller propagation delay than the NLFSR.

\section{Conditions for Invertibility} \label{main}

As we mentioned in Section~\ref{prev}, 
it is well-known how to construct invertible NLFSRs~\cite{Golomb_book}.
An $n$-bit NLFSR is invertible if, and only if, its
feedback function is of type:
\begin{equation} \label{eq1}
f(x_0,x_1,\ldots,x_{n-1}) = x_0 \oplus g(x_1,x_2,\ldots,x_{n-1}).
\end{equation}
The proof is simple, because every two consecutive states 
of an NLFSR overlap in $n-1$ positions.
This implies that each state can have only two possible 
predecessors and two possible successors. 
If $f$ is in the form (\ref{eq1}), then the NLFSR states which correspond to the binary $n$-tuples $x = (x_0, x_1, \ldots, x_{n-1})$
and $y = (\overline{x}_0, x_1, \ldots, x_{n-1})$
always have different successors. The values of $f(x)$ and $f(y)$
depend on the value of $g(x_1, \ldots, x_{n-1})$ and on the value of $x_0$. 
The value of $g(x_1, \ldots, x_{n-1})$ is the same for $x$ and $y$. 
The value of $x_0$ is different for $x$ and $y$.
Thus, $f(x) \not= f(y)$.
It is also easy to see that, if both $x$ and $y$ have the same successor, $f$
cannot have the form~(\ref{eq1}).

In the general case of mappings $\{0,1\}^n \rightarrow \{0,1\}^n$,
any binary $n$-tuple can have $2^n$ possible predecessors 
and $2^n$ possible successors. Therefore, to guarantee that 
a mapping $\{0,1\}^n \rightarrow \{0,1\}^n$ is 
invertible, we have to check that, for all $x, y \in \{0,1\}^n$, $x \not= y$ implies that $f_i(x) \not= f_i(y)$,
for some
$i \in \{0,1,\ldots,n-1\}$. This clearly requires 
the number of steps which is exponential in $n$.
The main contribution of this paper is a
more restricted sufficient condition
which can be checked in $O(n^2 N)$ steps.
To formulate this condition, we first introduce the  notion of a free variable of a function.

\begin{definition}[Free variable]
A variable $x_i \in dep(f)$ is called a free variable if
$f$ can decomposed as
\[
f(x_0,x_1,\ldots,x_{n-1}) = x_i \oplus g(x_0,x_1,\ldots,x_{n-1}).
\]
where $x_i \not\in dep(g_i)$.

\end{definition}

\begin{definition}[Set of free variables]
The set of free variables of $f$, $\Phi(f)$, contains all free variables of $f$
\[
\Phi(f) = \{x_i ~|~ x_i ~\mbox{is a free variable of}~ f\}.
\]
\end{definition}

Now we are ready to formulate the following sufficient condition for invertibility.

\begin{theorem} \label{t2}
A mapping of type~(\ref{map}) is invertible if the following two conditions hold:
\begin{enumerate}
\item For each $i \in \{0,1,\ldots,n-1\}$, $f_i$ is of type
\begin{equation} \label{th_main}
f_i(x_0,x_1,\ldots,x_{n-1}) = x_{j_i} \oplus g_i(x_0,x_1,\ldots,x_{n-1}),
\end{equation}
where $x_{j_i} \in \Phi(f_i)$.
\item Functions $f_{0}, f_{1}, \ldots, f_{n-1}$ can be re-ordered as $f_{i_0}, f_{i_1}, \ldots, f_{i_{n-1}}$ to satisfy the property
\[
dep(g_{i_k})  \subseteq  \left\{
\begin{array}{l}
 \emptyset, ~ \mbox{for} ~ k = 0\\
 \displaystyle\bigcup_{j=0}^{k-1} dep(f_{i_j}), ~ \mbox{for} ~ 1 \leq k \leq n-1 \\
\end{array}
\right.
\]
where ``$\cup$'' stands for the union.
\end{enumerate}
\end{theorem}
{\bf Proof:} By contradiction. 

Suppose that there exist a non-invertible mapping of type~(\ref{map}) for which the conditions of the theorem hold. Then, this mapping is of type
\begin{equation} \label{pr}
\left(
\begin{array}{l}
{x}_{i_0} \\
{x}_{i_1} \\
{x}_{i_2} \\
\ldots \\
{x}_{i_{n-1}} \\
\end{array}
\right)
\rightarrow
\left(
\begin{array}{l}
x_{j_0} \oplus g_{i_0} \\
x_{j_1} \oplus g_{i_1}(x_{j_0}) \\
x_{j_2} \oplus g_{i_2}(x_{j_0},x_{j_1}) \\
\ldots \\
x_{j_{n-1}} \oplus g_{i_{n-1}}(x_{j_0},x_{j_1}\ldots,x_{j_{n-2}}) \\
\end{array}
\right)
\end{equation}
where $x_{j_k} \in \Phi(f_{i_k})$, for $k \in \{0,1,\ldots,n-1\}$.

Since the mapping~(\ref{pr}) in non-invertible, there exist at least two input assignments $a = (a_{i_0},a_{i_1},\ldots,a_{i_{n-1}}) \in \{0,1\}^n$ and $a' = (a'_{i_0},a'_{i_1},\ldots,a'_{i_{n-1}}) \in \{0,1\}^n$, $a \not= a'$, which are mapped into the same output. We analyzing~(\ref{pr}) row by row, we can conclude that:
\begin{itemize}
\item ${a}_{i_1} \rightarrow a_{j_1} \oplus g_{i_0}$ and ${a'}_{i_0} \rightarrow a'_{j_0} \oplus g_{i_0}$ and $a_{j_0} \oplus g_{i_0} = a'_{j_0} \oplus g_{i_0}$ implies $a_{j_0} = a'_{j_0}$
\item $a_{i_1} \rightarrow a_{j_1} \oplus g_{i_1}(a_{j_0})$ and $a'_{i_1} \rightarrow a'_{j_1} \oplus g_{i_1}(a'_{j_0})$ and $a_{j_1} \oplus g_{i_1}(a_{j_0}) = a'_{j_1} \oplus g_{i_1}(a'_{j_0})$ and $a_{j_0} = a'_{j_0}$ implies $a_{j_1} = a'_{j_1}$
\item $\ldots$
\item 
$a_{i_{n-1}} \rightarrow a_{j_{n-1}} \oplus g_{i_{n-1}}(a_{j_0},a_{j_1}\ldots,a_{j_{n-2}})$ 
and 
$a'_{i_{n-1}} \rightarrow a'_{j_{n-1}} \oplus g_{i_{n-1}}(a'_{j_0},a'_{j_1}\ldots,a'_{j_{n-2}})$ 
and 
$a_{j_{n-1}} \oplus g_{i_{n-1}}(a_{j_0},a_{j_1}\ldots,a_{j_{n-2}}) = a'_{j_{n-1}} \oplus g_{i_{n-1}}(a'_{j_0},a'_{j_1}\ldots,a'_{j_{n-2}})$ 
and
$a_{j_0} = a'_{j_0}$, $a_{j_1} = a'_{j_1}$, $\ldots$ $a_{j_{n-2}} = a'_{j_{n-2}}$
implies $a_{j_{n-1}} = a'_{j_{n-1}}$.
\end{itemize}
Therefore, $a = a'$. We reached a contradiction. Thus, the assumption that there exist a non-invertible mapping of type~(\ref{map}) for which the conditions of the theorem hold is not true.
\begin{flushright}
$\Box$
\end{flushright}

An example of mapping which satisfies the conditions of Theorem~\ref{t2} is: 
\begin{equation} \label{cons}
\left(
\begin{array}{l}
{x}_0 \\
{x}_1 \\
{x}_2 \\
\ldots \\
{x}_{n-1} \\
\end{array}
\right)
\rightarrow
\left(
\begin{array}{l}
x_1 \oplus g_0 \\
x_2 \oplus g_1(x_1) \\
x_3 \oplus g_2(x_1,x_2) \\
\ldots \\
x_0 \oplus g_{n-1}(x_1,x_2\ldots,x_{n-1}) \\
\end{array}
\right)
\end{equation}

The reader may notice that mappings defined by Theorem~\ref{t2}
can be obtained by re-labeling variables in a T-function.
The re-labeling is given by the ordering of free variables.
So, the mapping~(\ref{th_main}) can be viewed
as a composition of a bit permutation and a T-function.
Since there are $n!$ bit permutations, the class of invertible mappings considered in this paper is
by a factor of $n!$ larger than the class of invertible mappings 
based on T-functions.

\begin{figure}[t!]
\begin{center}
\includegraphics*[angle=270,width=1.6in]{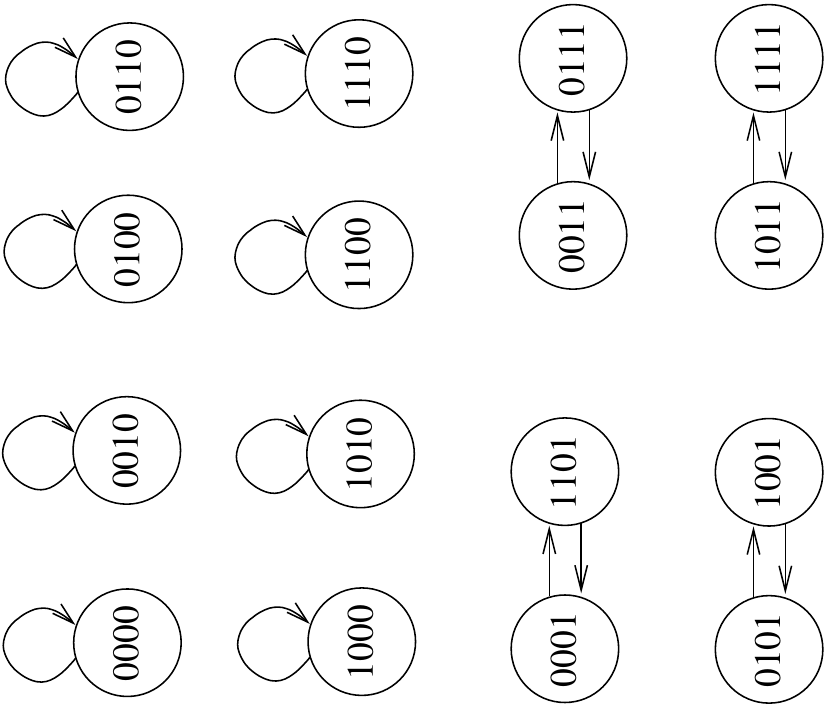}
\caption{The state transition graph of the mapping~(\ref{m1}).}\label{fm1}
\end{center}
\end{figure}

\begin{figure}[t!]
\begin{center}
\includegraphics*[angle=270,width=2.7in]{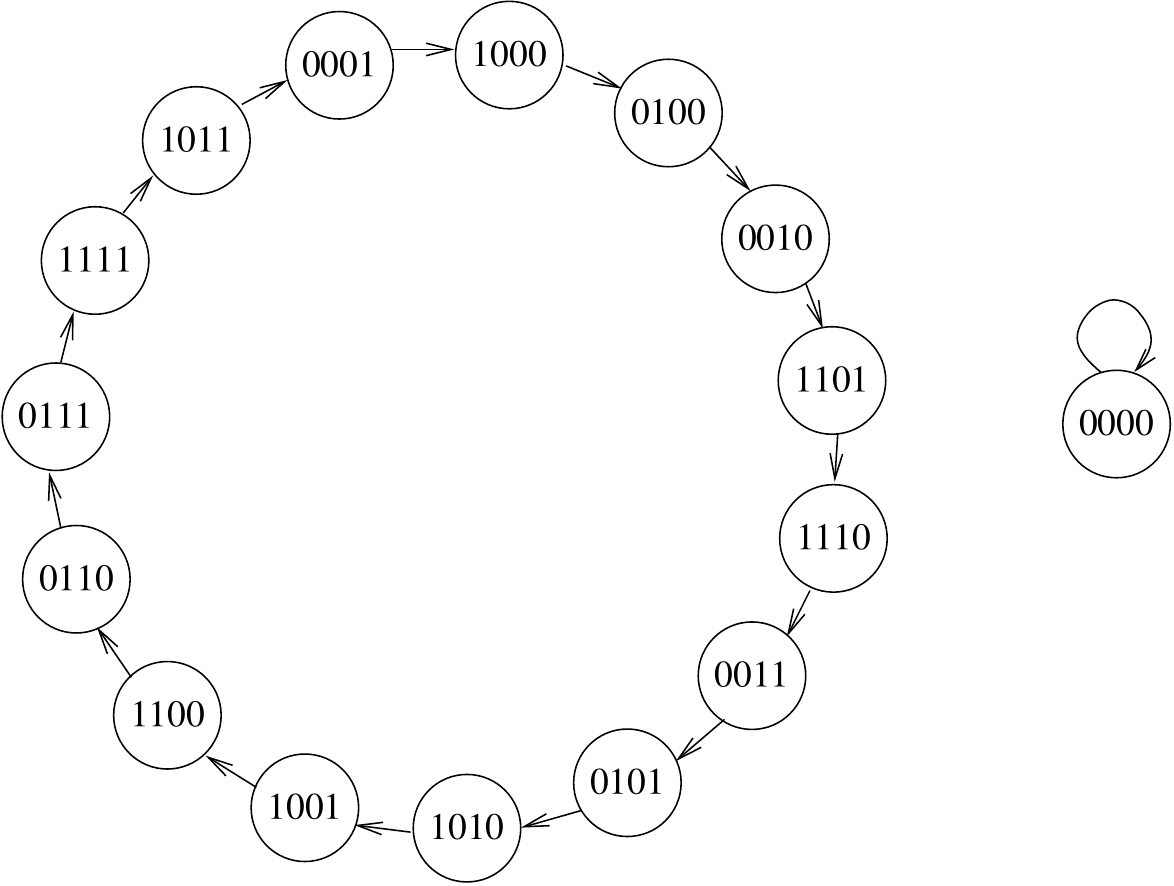}
\caption{The state transition graph of the mapping~(\ref{m2}).}\label{fm2}
\end{center}
\end{figure}

Clearly, the re-labeling of variables does not change the implementation cost of a mapping.
However, it might drastically change the cycle structure of the underlying 
state transition graph, as illustrated by the following example. 
Consider the following 4-variable mapping based on a T-function:
\begin{equation} \label{m1}
\left(
\begin{array}{c}
x_0\\
x_1\\
x_2\\
x_3\\
\end{array}
\right)
\rightarrow
\left(
\begin{array}{c}
x_0\\
x_1\\
x_2 \oplus x_0\\
x_3 \oplus x_0 \oplus x_0 x_1\\
\end{array}
\right).
\end{equation}
This mapping has a quite uninteresting state transition graph shown in Figure~\ref{fm1}.
Let us re-label the variables as $(0,1,2,3) \rightarrow (1,2,3,0)$. We get the mapping:
\begin{equation} \label{m2}
\left(
\begin{array}{c}
x_0\\
x_1\\
x_2\\
x_3\\
\end{array}
\right)
\rightarrow
\left(
\begin{array}{c}
x_1\\
x_2\\
x_3 \oplus x_1\\
x_0 \oplus x_1 \oplus x_1 x_2\\
\end{array}
\right).
\end{equation}
which has the state transition graph shown in Figure~\ref{fm2}. All states, except the all-0 state, are included in one cycle. If we implement this mapping by the binary machine shown in Figure~\ref{f2} and initialize the binary machine to any non-zero state, then its output generates a pseudo-random sequence with period 15 which 
satisfies the first two randomness postulates of Golomb~\cite{Golomb_book}, namely the balance property and the run property. 
No sequence generated by a nonlinear Boolean function satisfy the third randomness postulate (two-level autocorrelation property). 
Therefore, the mapping we obtained by re-labeling has much more interesting statistical properties than the original mapping. 

The cycle structure of a mapping is important for some cryptographic applications, e.g. stream ciphers~\cite{robshaw94stream}. Obviously, if we iterate a mapping a large number of times, we do not want the sequence of generated states to be trapped in a short cycle. 
It is worth noting 
that formal verification tools based on reachability analysis can be adopted for finding short cycles in a state transition graph. There are dedicated tools to compute the number and length of cycles in state transition graphs of synchronous sequential networks, e.g. BNS~\cite{DuT11} which is based on SAT-based bounded model checking and BooleNet~\cite{DuTM05} which is based on Binary Decision Diagrams. 

The mappings defined by Theorem~\ref{t2} are invertible for any choice of Boolean functions $g_i(x_0,x_2\ldots,x_{i-1})$. 
The smaller is the number of Boolean operations in the expressions of $g_i$'s,
the smaller is its hardware implementation cost.
Note, that we are not restricted to represent $g_i$'s in ANF. Any Boolean expression combining AND, OR, NOT and XOR can be used. 
Multi-level logic optimization tools, such as UC Berkeley tool ABC~\cite{abc} 
can be applied to transform the ANF or other Boolean expression representing the function into 
an optimized multi-level expression.
Clearly, the choice of $g_i$'s will be guided not only by the hardware cost,
but also by other criteria determining the cryptographic strength of the resulting function, e.g. nonlinearity, correlation immunity, algebraic degree, etc. (see~\cite{CuS09} for requirements on cryptographically strong functions).

Finally, we would like to point out that the conditions of Theorem~\ref{t2} are sufficient, but not necessary conditions for invertibility. For example, the following 4-variable mapping is invertible, but it does not satisfy them:
\[
\left(
\begin{array}{c}
x_0\\
x_1\\
x_2\\
x_3\\
\end{array}
\right)
\rightarrow
\left(
\begin{array}{c}
x_1 \oplus x_0 \oplus x_0 x_2\\
x_2 \oplus x_1 \oplus x_3\\
x_3 \oplus x_0 x_2\\
x_0\\
\end{array}
\right).
\]
%It has two cycles:
%a cycle consisting of 15 states $(0001, 0110,$ $1000, 1001, 1111, 1101, 0011, 0010, 0100, 1100, 
%0101, 1010, 0111, 1110, 1011)$ and a cycle consisting of an all-zero state with a self-loop.

\section{Condition Checking} \label{compl}

Next, let us estimate the number of steps required to check the conditions of Theorem~\ref{t2}.
Suppose that all Boolean functions of the mapping are represented in ANF. 
Let $N$ be the ANF size of the largest function in the mapping. 
The {\em size} of an ANF is defined as the total number of variables appearing in the ANF. 
For example, the ANF $x_1 \oplus x_1 x_2$ is of size 3.

\begin{algorithm}[t!]
\caption{\footnotesize{Checks conditions of Theorem~\ref{t2} for a mapping of type~(\ref{map})
in which all Boolean functions are represented in ANF}} \label{alg}
\begin{algorithmic}[1]
\STATE $\Phi = \emptyset$; 
\STATE flag = 0;
\FOR{every $i$ from 0 to $n-1$} 
\IF{$f_i = x_j$ or $f_i = x_j \oplus 1$, for some $j \in \{0,1,\ldots,n-1\}$} 
\STATE add $x_j$ to $\Phi$ 
\STATE mark $f_i$ 
\STATE flag = 1; 
\ENDIF
\ENDFOR
\IF{flag = 0} 
\STATE Return "Conditions are not satisfied"
\ENDIF
\STATE {\bf loop:} 
\FOR{every $i$ from 0 to $n-1$} 
\IF{$f_i$ is not marked} 
\IF{$f_i$ is of type~(\ref{th_main}) and $x_k \in \Phi~ \forall x_k \in dep(g_i)$} 
\STATE add $x_j$ to $\Phi$ 
\STATE mark $f_i$
\STATE go to {\bf loop}
\ENDIF
\ENDIF
\ENDFOR
\FOR{every $i$ from 0 to $n-1$} 
\IF{$f_i$ is marked} 
\STATE continue
\ELSE
\STATE Return "Conditions are not satisfied"
\ENDIF
\ENDFOR
\STATE Return "Mapping is invertible"
\end{algorithmic}
\end{algorithm}

Consider the pseudocode shown as Algorithm~\ref{alg}.
$\Phi$ is a set which keeps track of variables which are identified as free for some $f_i$. If this set is implemented as a hash table of size $n$, then adding a variable to the set or checking if a variable belongs to the set can be done in constant time.

In the first {\bf for}-loop, we check if each $f_i$ is of type $f_i = x_j$
or $f_i = x_j \oplus 1$, for some $j \in \{0,1,\ldots,n-1\}$. If yes, we 
add $x_j$ to the set $\Phi$ and mark $f_i$. Since the steps 4, 5, 6 and 7 can be done 
in $O(1)$ time, the complexity of the first {\bf for}-loop is $O(n)$.

If none of the functions $f_i$ are marked during the first {\bf for}-loop, the algorithm terminates with the conclusion that the conditions of Theorem~\ref{t2} are not satisfied.

In the second {\bf for}-loop, for each non-marked $f_i$, we check if it is of type~(\ref{th_main}) 
and if every $x_k \in dep(g_i)$ belongs to $\Phi$.
If yes, add $x_j$ to $\Phi$, mark $f_i$, and return to step 13.
The steps 15, 17 and 18 can be done in $O(1)$ time.
The step 16 requires $O(N)$ time.
Thus, the complexity of the second {\bf for}-loop is $O(n N)$.
Since we return to the step 13 at most $n$ times, 
the overall complexity of steps 13-22 is $O(n^2 N)$.

In the third {\bf for}-loop, we check if all $f_i$ are marked.
If yes, the mapping is invertible. Otherwise, the algorithm
returns "conditions are not satisfied". Since the
conditions of Theorem~\ref{t2} are sufficient, but not necessary conditions for invertibility, in this case the mapping may or may not be invertible. The complexity of the third {\bf for}-loop is $O(n)$.
We can conclude that overall complexity of Algorithm~\ref{alg} is $O(n^2 N)$.

\section{Applications} \label{appl}

In this section we show how the presented results can be used in stream cipher design.

A possible way to construct a key stream generator for a stream cipher is  
to run several FSRs in parallel and to combine their outputs with a nonlinear Boolean function~\cite{CuS09}. 
The resulting structure is called a {\em combination generator}.
If the periods of FSRs are pairwise co-prime, then the period of the resulting key stream generator
is equal to the product of periods of FRSs~\cite{DuT05}. Examples of stream ciphers based on combination generators are VEST~\cite{cryptoeprint:2005:415}, Achterbahn-128/80~\cite{GaGK07}, and the cipher~\cite{GaGK06}. VEST uses 16 10-bit and 16 11-bit NLFSRs. Achterbahn-128/80 uses 13 NLFSRs of size from 21 to 33 bits. The cipher from~\cite{GaGK06} uses 10 NLFSRs of size 22-29, 31 and 32 bits.

At present it is not known how to construct large NLFSRs with a guaranteed long period.
Existing algorithms cover special cases only, e.g.~\cite{Du12a,Du13a}. 
Small NLFSRs which are used in combination generators are computed by a random search.
Lists of $n$-bit NLFSRs of size $< 25$ bits with the period $2^n-1$ whose feedback functions contain up to 6 binary Boolean operations in their ANFs are available in~\cite{Du12l}. 
It is known that, for example, there are no 20-bit NLFSRs with the period $2^{20}-1$ whose feedback function contains only 
four binary Boolean operations in its ANF (i.e. implementable with four 2-input gates).

Using Algorithm~\ref{alg} to bound random search, in 12 hours we were able to find 63 20-variable nonlinear invertible mappings with the period $2^{20}-1$ whose functions $f_i$ contain no more than 4 binary Boolean operations in their ANFs in total. Two representatives are:

\begin{enumerate}
\item  $f_{16} = x_{17} \oplus x_4 x_{11}$, $f_{13} = x_{14} \oplus x_{13}$, $f_4 = x_5 \oplus x_1$. \vspace{1mm}
\item  $f_{19} = x_{0} \oplus x_{16}$, $f_{15} = x_{16} \oplus x_{3} \oplus x_1 x_{15}$.
\end{enumerate}
The omitted functions are of type $f_i = x_{(i+1)~mod~20}$, for $i \in \{0,1,\ldots,19\}$. 
This shows that Algorithm~\ref{alg} is useful for finding nonlinear invertible 
mappings with a small hardware implementation cost.
Other important properties, such as nonlinearity, correlation immunity, algebraic degree, etc., can then be used to further guide the search.

\section{Conclusion} \label{conc}

We derived a sufficient condition for invertibility of mappings of type $\{0,1,\ldots,2^n-1\} \rightarrow \{0,1,\ldots,2^n-1\}$ which can be checked in $O(n^2 N)$ steps, where $N$ 
is the ANF size of the largest Boolean function in the mapping.
The presented method can be used as a starting point for
constructing nonlinear invertible mappings which can be efficiently implemented in hardware.

Future work remains on constructing new cryptographic primitives based on the presented class of invertible mappings and evaluating their security and hardware cost. We also plan to investigate the usability our results in reversible computing.

\section{Acknowledgements}
This work was supported in part by the research grant No SM14-0016 from the Swedish Foundation for Strategic Research. The author would like to thank the anonymous reviewers for their valuable comments and suggestions to improve the
quality of the paper.

%\bibliographystyle{ieeetr}
%\bibliography{bib}

\begin{thebibliography}{10}

\bibitem{Cy14}
{Center for Strategic and International Studies}, ``Net losses: Estimating the
  global cost of cybercrime,'' June 2014.
\newblock https://www.mcafee.com/mx/resources/
  reports/rp-economic-impact-cybercrime2.pdf.

\bibitem{Er12}
{Ericsson}, ``More that 50 billions connected devices,'' 2012.
\newblock www.ericsson.com/res/docs/whitepapers/wp-50-billions.pdf.

\bibitem{Pr14}
{Proofpoint}, ``Proofpoint uncovers {I}nternet of {T}hings ({IoT})
  cyberattack,'' January 2014.
\newblock https://www.proofpoint.com/us/
  proofpoint-uncovers-internet-of-things-iot-cyberattack.

\bibitem{Er5G}
{Ericsson}, ``5g security: Scenarios and solutions,'' 2015.
\newblock www.ericsson.com/news/150624-wp-5g-security\_244069646\_c.

\bibitem{St06}
D.~Stinson, {\em Cryptography Theory and Practice}.
\newblock Chapman \& Hall/CRC, 3rd edition, 2006.

\bibitem{pres07}
A.~Bogdanov, L.~Knudsen, G.~Leander, C.~Paar, A.~Poschmann, M.~Robshaw,
  Y.~Seurin, and C.~Vikkelsoe, ``Present: An ultra-lightweight block cipher,''
  in {\em Cryptographic Hardware and Embedded Systems - CHES 2007} (P.~Paillier
  and I.~Verbauwhede, eds.), vol.~4727 of {\em Lecture Notes in Computer
  Science}, pp.~450--466, Springer Berlin Heidelberg, 2007.

\bibitem{pri12}
J.~Borghoff, A.~Canteaut, T.~Güneysu, E.~Kavun, M.~Knezevic, L.~Knudsen,
  G.~Leander, V.~Nikov, C.~Paar, C.~Rechberger, P.~Rombouts, S.~Thomsen, and
  T.~Yalçın, ``Prince – a low-latency block cipher for pervasive computing
  applications,'' in {\em Advances in Cryptology – ASIACRYPT 2012} (X.~Wang
  and K.~Sako, eds.), vol.~7658 of {\em Lecture Notes in Computer Science},
  pp.~208--225, Springer Berlin Heidelberg, 2012.

\bibitem{hell-grain}
M.~Hell, T.~Johansson, A.~Maximov, and W.~Meier, ``The {G}rain family of stream
  ciphers,'' {\em New Stream Cipher Designs: The eSTREAM Finalists, LNCS 4986},
  pp.~179--190, 2008.

\bibitem{DuH14}
E.~Dubrova and M.~Hell, ``Espresso: A stream cipher for {5G} wireless
  communication systems,'' {\em Cryptography and Communications}, 2015.
\newblock accepted, availible at https://eprint.iacr.org/2015/241.

\bibitem{Quark}
J.-P. Aumasson, L.~Henzen, W.~Meier, and M.~Naya-Plasencia, ``Quark: A
  lightweight hash,'' {\em Journal of Cryptology}, vol.~26, no.~2,
  pp.~313--339, 2013.

\bibitem{DuNS15}
E.~Dubrova, M.~Naslund, and G.~Selander, ``{CRC}-based message authentication
  for {5G} mobile technology,'' in {\em Proceedings of 1st IEEE International
  Workshop on 5G Security}, August 2015.

\bibitem{CuS09}
T.~W. Cusick and P.~St{\v a}nic{\v a}, {\em Cryptographic {B}oolean functions
  and applications}.
\newblock San Diego, CA, USA: Academic Press, 2009.

\bibitem{Gr91}
D.~H. Green, ``Families of {R}eed-{M}uller canonical forms,'' {\em
  International Journal of Electronics}, vol.~70, pp.~259--280, 1991.

\bibitem{espr}
R.~K. Brayton, C.~McMullen, G.~Hatchel, and A.~Sangiovanni-Vincentelli, {\em
  Logic Minimization Algorithms For VLSI Synthesis}.
\newblock Kluwer Academic Publishers, 1984.

\bibitem{Fe73}
H.~Feistel, ``Cryptography and computer privacy,'' {\em Scientific American},
  vol.~228, pp.~15--23, May 1973.

\bibitem{DES77}
{National Bureau of Standards}, ``Data encryption standard,'' Tech. Rep. NBS
  FIPS PUB 46, U.S. Department of Commerce, January 1977.

\bibitem{ScK96}
B.~Schneier and J.~Kelsey, ``Unbalanced feistel networks and block-cipher
  design,'' in {\em Fast Software Encryption, 3rd International Workshop
  Proceedings}, pp.~121--144, Springer-Verlag, 1996.

\bibitem{Sh93}
A.~Shamir, ``Efficient signature schemes based on birational permutations,'' in
  {\em Proceedings of the 13th Annual International Cryptology Conference on
  Advances in Cryptology}, CRYPTO'93, (London, UK, UK), pp.~1--12,
  Springer-Verlag, 1993.

\bibitem{KlS02}
A.~Klimov and A.~Shamir, ``A new class of invertible mappings,'' in {\em
  Revised Papers from the 4th International Workshop on Cryptographic Hardware
  and Embedded Systems}, CHES'02, (London, UK), pp.~470--483, Springer-Verlag,
  2002.

\bibitem{Kl05}
A.~Klimov, {\em Applications of T-functions in Cryptography}.
\newblock Ph.D. Thesis, Weizmann Institute of Science, 2005.

\bibitem{LiH94}
R.~Lidl and H.~Niederreiter, {\em Introduction to Finite Fields and their
  Applications}.
\newblock Cambridge Univ. Press, 1994.

\bibitem{Di01}
L.~E. Dickson, {\em Linear Groups with an Exposition of the Galois Field
  Theory}.
\newblock Teubner, 1901.

\bibitem{Ri99}
R.~L. Rivest, ``Permutation polynomials modulo $2^w$,'' {\em Finite Fields and
  Their Applications}, vol.~7, pp.~287--292, 1999.

\bibitem{Ja89}
C.~J.~A. Jansen, {\em Investigations On Nonlinear Streamcipher Systems:
  Construction and Evaluation Methods}.
\newblock Ph.D. Thesis, Technical University of Delft, 1989.

\bibitem{Golomb_book}
S.~Golomb, {\em Shift Register Sequences}.
\newblock Aegean Park Press, 1982.

\bibitem{LiD14}
N.~Li and E.~Dubrova, ``An algorithm for constructing a smallest register with
  non-linear update generating a given binary sequence,'' in {\em Proceedings
  of IEEE International Symposium on Multiple-Valued Logic (ISMVL'2014)}, 2014.

\bibitem{Du10aj}
E.~Dubrova, ``Synthesis of binary machines,'' {\em IEEE Transactions on
  Information Theory}, vol.~57, pp.~6890 -- 6893, 2011.

\bibitem{Du11a}
E.~Dubrova, ``Synthesis of parallel binary machines,'' in {\em Proceedings of
  International Conference of Computer-Aided Design (ICCAD'2011)}, (San Jose,
  CA, USA), pp.~200--206, Nov. 2011.

\bibitem{robshaw94stream}
M.~Robshaw, ``Stream ciphers,'' Tech. Rep. TR - 701, July 1994.

\bibitem{DuT11}
E.~Dubrova and M.~Teslenko, ``A {SAT}-based algorithm for finding attractors in
  synchronous {B}oolean networks,'' {\em IEEE/ACM Transactions on Computational
  Biology and Bioinformatics}, vol.~8, no.~5, pp.~1393 --1399, 2011.

\bibitem{DuTM05}
E.~Dubrova, M.~Teslenko, and A.~Martinelli, ``Kauffman networks: analysis and
  applications,'' in {\em IEEE/ACM International Conference on Computer-Aided
  Design (ICCAD'2005)}, pp.~479--484, Nov 2005.

\bibitem{abc}
{Berkeley Logic Synthesis and Verification Group}, ``{ABC}: A system for
  sequential synthesis and verification, release 70930,'' 2007.

\bibitem{DuT05}
E.~Dubrova and M.~Teslenko, ``Compositional properties of {R}andom {B}oolean
  {N}etworks,'' {\em Physical Review E}, vol.~71, p.~056116, May 2005.

\bibitem{cryptoeprint:2005:415}
B.~Gittins, H.~A. Landman, S.~O'Neil, and R.~Kelson, ``A presentation on {VEST}
  hardware performance, chip area measurements, power consumption estimates and
  benchmarking in relation to the aes, sha-256 and sha-512.'' Cryptology ePrint
  Archive, Report 2005/415, 2005.
\newblock http://eprint.iacr.org/2005/415.

\bibitem{GaGK07}
B.~Gammel, R.~G{\"o}ttfert, and O.~Kniffler, ``Achterbahn-128/80: Design and
  analysis,'' in {\em SASC'2007: Workshop Record of The State of the Art of
  Stream Ciphers}, pp.~152--165, 2007.

\bibitem{GaGK06}
B.~M. Gammel, R.~G{\"o}ttfert, and O.~Kniffler, ``An {NLFSR}-based stream
  cipher,'' in {\em ISCAS}, 2006.

\bibitem{Du12a}
E.~Dubrova, ``A method for generating full cycles by a composition of
  {NLFSR}s,'' {\em Design, Codes and Cryptography}, 2012.

\bibitem{Du13a}
E.~Dubrova, ``A scalable method for constructing {G}alois {NLFSR}s with period
  $2^n-1$ using cross-join pairs,'' {\em IEEE Transactions on Information
  Theory}, vol.~1, no.~59, pp.~703--709, 2013.

\bibitem{Du12l}
E.~Dubrova, ``A list of maximum-period {NLFSR}s.'' Cryptology ePrint Archive,
  Report 2012/166, 2012.
\newblock http://eprint.iacr.org/2012/166.

\end{thebibliography}

\end{document}